\documentclass[journal]{IEEEtran}
\usepackage{amsmath,amssymb}
\usepackage{epsf}
\usepackage{graphicx}
\usepackage{subfigure}
\usepackage{balance}
\usepackage[sort,compress]{cite}
\usepackage{multicol}
\usepackage{nccmath}
\usepackage[linesnumbered,ruled,slide]{algorithm2e}
\usepackage{color}
\usepackage[normalem]{ulem}
\usepackage{multirow}
\usepackage{amsthm}
\usepackage{ulem}
\usepackage{slashbox}
\usepackage{longtable}
\usepackage{multirow}
\usepackage{rotating}

\begin{document}

\title{DeepRMSA: A Deep Reinforcement Learning Framework for Routing, Modulation and Spectrum Assignment in Elastic Optical Networks}

\author{~Xiaoliang~Chen,~\IEEEmembership{Member,~IEEE},~Baojia~Li,~Roberto~Proietti,~Hongbo~Lu,~Zuqing~Zhu,~\IEEEmembership{Senior Member,~IEEE},~S.~J.~Ben~Yoo,~\IEEEmembership{Fellow,~IEEE,~Fellow,~OSA}

\thanks{X. Chen, R. Proietti, H. Lu and S. J. B. Yoo are with the Department of Electrical and Computer Engineering, University of California, Davis, Davis, CA 95616, USA (Email: xlichen@ucdavis.edu, sbyoo@ucdavis.edu).}
\thanks{B. Li and Z. Zhu are with the School of Information Science and Technology, University of Science and Technology of China, Hefei, Anhui 230027, P. R. China (Email: zqzhu@ieee.org).}
\thanks{Manuscript received Dec. 8, 2018.}}

\maketitle

\begin{abstract}
This paper proposes DeepRMSA, a deep reinforcement learning framework for routing, modulation and spectrum assignment (RMSA) in elastic optical networks (EONs). DeepRMSA learns the correct online RMSA policies by parameterizing the policies with deep neural networks (DNNs) that can sense complex EON states. The DNNs are trained with experiences of dynamic lightpath provisioning. We first modify the asynchronous advantage actor-critic algorithm and present an episode-based training mechanism for DeepRMSA, namely, DeepRMSA-EP. DeepRMSA-EP divides the dynamic provisioning process into multiple episodes (each containing the servicing of a fixed number of lightpath requests) and performs training by the end of each episode. The optimization target of DeepRMSA-EP at each step of servicing a request is to maximize the cumulative reward within the rest of the episode. Thus, we obviate the need for estimating the rewards related to unknown future states. To overcome the instability issue in the training of DeepRMSA-EP due to the oscillations of cumulative rewards, we further propose a window-based flexible training mechanism, i.e., DeepRMSA-FLX. DeepRMSA-FLX attempts to smooth out the oscillations by defining the optimization scope at each step as a sliding window, and ensuring that the cumulative rewards always include rewards from a fixed number of requests. Evaluations with the two sample topologies show that DeepRMSA-FLX can effectively stabilize the training while achieving blocking probability reductions of more than $20.3\%$ and $14.3\%$, when compared with the baselines.
\end{abstract}

\IEEEpeerreviewmaketitle

\section{Introduction}
\IEEEPARstart{T}{he} explosive growth of emerging applications (e.g., cloud computing) and the popular adoption of new networking paradigms (e.g., the Internet of Things) are demanding a new network infrastructure that can support dynamic, high-capacity and quality-of-transmission (QoT)-guaranteed end-to-end services. Recently, elastic optical networking (EON) has emerged as one of the most promising networking technologies for the next-generation backbone networks \cite{Gerstel2012}. Compared with the traditional fixed-grid (e.g., $50$ GHz) wavelength-division multiplexing (WDM) scheme, EON can flexibly set up bandwidth-variable superchannels by grooming series of finer-granularity (e.g., $6.25$ GHz) subcarriers and adapting the modulation formats according to the QoT of lightpaths \cite{Jinno2010}.

The flexible resource allocation mechanisms in EON, on the other hand, make the corresponding service provisioning designs more complicated. To fully exploit the benefits of such flexibilities and realize cost-effective EON, previous studies have intensively investigated the routing, modulation and spectrum assignment (RMSA) problem for EON \cite{RMSA_survey}. The authors of \cite{Wang2011,Christodoulopoulos2011,Klinkowski2013} first proposed integer linear programming (ILP) models for solving the static RMSA problems, where all the lightpath requests are assumed to be known in prior. While the ILP models can provide the optimal solutions to the RMSA problems, they are proved to be $\mathcal{NP}$-hard \cite{Wang2011} and are intractable for large-scale problems. In this context, a number of heuristic or approximation algorithms have been developed. In \cite{Wang2011}, Wang \emph{et al.} proposed two algorithms, namely, balanced load spectrum allocation and shortest path with maximum spectrum reuse, to minimize the maximum required spectrum resources in an EON accounting for the given traffic demand. The authors of \cite{Christodoulopoulos2011} presented a simulated annealing approach for determining the servicing order of lightpath requests and applied the k-shortest path routing and first-fit (KSP-FF) scheme to calculate the RMSA solution for each request afterward. In \cite{Gong2012,Klinkowski2013}, the authors investigated to leverage genetic algorithms to realize joint RMSA optimizations. A conflict graph based two-phase algorithm with proved performance level was proposed in \cite{Wu2017}. For more heuristic RMSA designs, such as random-fit, exact-fit and most-used spectrum assignment, readers can refer to \cite{RMSA_survey}.

Unlike static RMSA problems for which explicit optimization models can be formulated, optimizing dynamic lightpath provisioning in EONs (i.e., dynamic RMSA problems) is more challenging. The dynamic arrivals and departures of lightpath requests as well as the uncertainty of future traffic could dramatically destabilize the EON state and thus deteriorate the efficiency of the optimizations based on the current state. To cope with such dynamics, a few dynamic RMSA designs have been reported lately, in addition to those that can be derived from the aforementioned static RMSA algorithms. The authors in \cite{Zhu2013} applied the multi-path routing scheme and developed several empirical weighting methods taking into account path lengths, link spectrum utilization, and other features to realize state-aware dynamic RMSA. In \cite{Yin2013}, Yin \emph{et al.} investigated the spectrum fragmentation effect in dynamic lightpath provisioning and proposed a fragmentation-aware RMSA algorithm to mitigate spectrum fragmentation. More aggressive service reconfiguration approaches, e.g., spectrum defragmentation \cite{Cugini13,Zhang2016}, have also been proposed as complements to normal RMSA algorithms to enable periodical service consolidations but at the expense of high operational costs. However, the existing works only apply fixed RMSA policies regardless of the time-varying EON states or rely on simple empirical policies based on manually extracted features, i.e., lack of comprehensive perceptions of the holistic EON states, and therefore are unable to achieve real adaptive service provisioning in EONs.

In the meantime, recent advances in deep reinforcement learning (DRL) have demonstrated beyond human-level performance in handling large-scale online control tasks \cite{humanlevel,A3C}. By parameterizing policies with deep neural networks (DNNs) \cite{lecun2015}, DRL enables learning agents to perceive complex system states from high-dimensional input data (e.g., screenshots and traffic matrices) and progressively learn correct policies through experiences of repeated interactions with the target systems. The application of DRL in the communication and networking domain has received intensive research interests during the past two years \cite{Luong2018,Xu2018,Salman2018}. In \cite{Xu2018}, the authors enhanced the general deep Q-learning framework in \cite{humanlevel} with novel exploration and experience replay techniques to solve the traffic engineering problem. The authors of \cite{Salman2018} presented a DRL-based framework for datacenter network management and demonstrated a DRL agent which can learn the optimal topology configurations with respect to different application profiles. Nevertheless, the application of DRL in optical networking, or in particular, for addressing the RMSA problem, has not been investigated.

In this paper, we propose DeepRMSA, a DRL-based RMSA framework for learning the optimal online RMSA policies in EONs. The contributions of this paper can be summarized as follows. 1) We propose, for the first time, a DRL framework for optical network management and resource allocation, i.e., RMSA. 2) We propose two training mechanisms for DeepRMSA, taking into account the unique characteristics of the RMSA problem. 3) Numerical results verify the superiority of DeepRMSA over the state-of-art heuristic algorithms.

The rest of the paper is organized as follows. Section \ref{sec:formulation} presents the RMSA problem formulation. Section \ref{sec:framework} details the DeepRMSA framework. In Section \ref{sec:design}, we elaborate on the design of DeepRMSA, including the modeling and the training mechanisms. Then, in Section \ref{sec:evaluation}, we show the performance evaluations and related discussions. Finally, Section \ref{sec:conclusion} concludes the paper.

\section{Problem Formulation}\label{sec:formulation}
Let $G(V,E,F)$ denote an EON topology, where $V$ and $E$ represent the sets of nodes and fiber links, $F = \left\{F_{e,f} \mid _{e,f}\right\}$ contains the state of each frequency slot (FS) $f \in [1,f_0]$ on each fiber link $e \in E$. We model a lightpath request from node $o$ to $d$ ($o, d \in V$) as $\mathcal{R}_t(o,d,b,\tau)$, with $b$ Gb/s and $\tau$ denoting the bandwidth requirement and service duration, respectively. To provision $\mathcal{R}_t$, we need to compute an end-to-end routing path $\mathcal{P}_{o,d}$, determine a proper modulation format $m$ to use for QoT assurance, and allocate a number of spectrally contiguous FS's (i.e., the spectrum contiguous constraint) on each link along $\mathcal{P}_{o,d}$ according to $b$ and $m$. In this work, we assume that the EON is not equipped with the spectrum conversion capability. Therefore, the spectra allocated on different fibers to $\mathcal{R}_t$ must align (i.e., the spectrum continuous constraint). We adopt the impairment-aware model in \cite{kozicki2010} to decide the modulation format according to the physical distance of $\mathcal{P}_{o,d}$. Specifically, the number of FS's needed can be computed as,
\begin{equation}\label{eq:1}
\small
n = \left \lceil \frac{b}{m \cdot C_{grid}^{BPSK}} \right \rceil,
\end{equation}
where $C_{grid}^{BPSK}$ is the data rate an FS of BPSK signal can support and $m \in \left[1,2,3,4\right]$ corresponds to BPSK, QPSK, 8-QAM and 16-QAM, respectively. The static RMSA problem (i.e., offline network planning) gives a set of permanent lightpath requests $\mathcal{R} = \left\{\mathcal{R}_{t} \mid _{t}\right\}$ ($\tau \rightarrow \infty$) and requires provisioning all of them in a batch following the link capacity constraint \cite{Wang2011}. The objective of the static RMSA problem is to minimize the total spectrum usage. Unlike the static problem where requests are known in prior, in the dynamic RMSA problem (i.e., online lightpath provisioning) being considered in this work, lightpath requests arrive and expire on-the-fly and need to be serviced immediately upon their arrivals. The dynamic RMSA problem aims at minimizing the long-term request blocking probability, which is defined as the ratio of the number of blocked requests to the total number of requests over a period.

\section{DeepRMSA Framework}\label{sec:framework}
\begin{figure}%[h]
\begin{center}
\includegraphics[width=8.5cm]{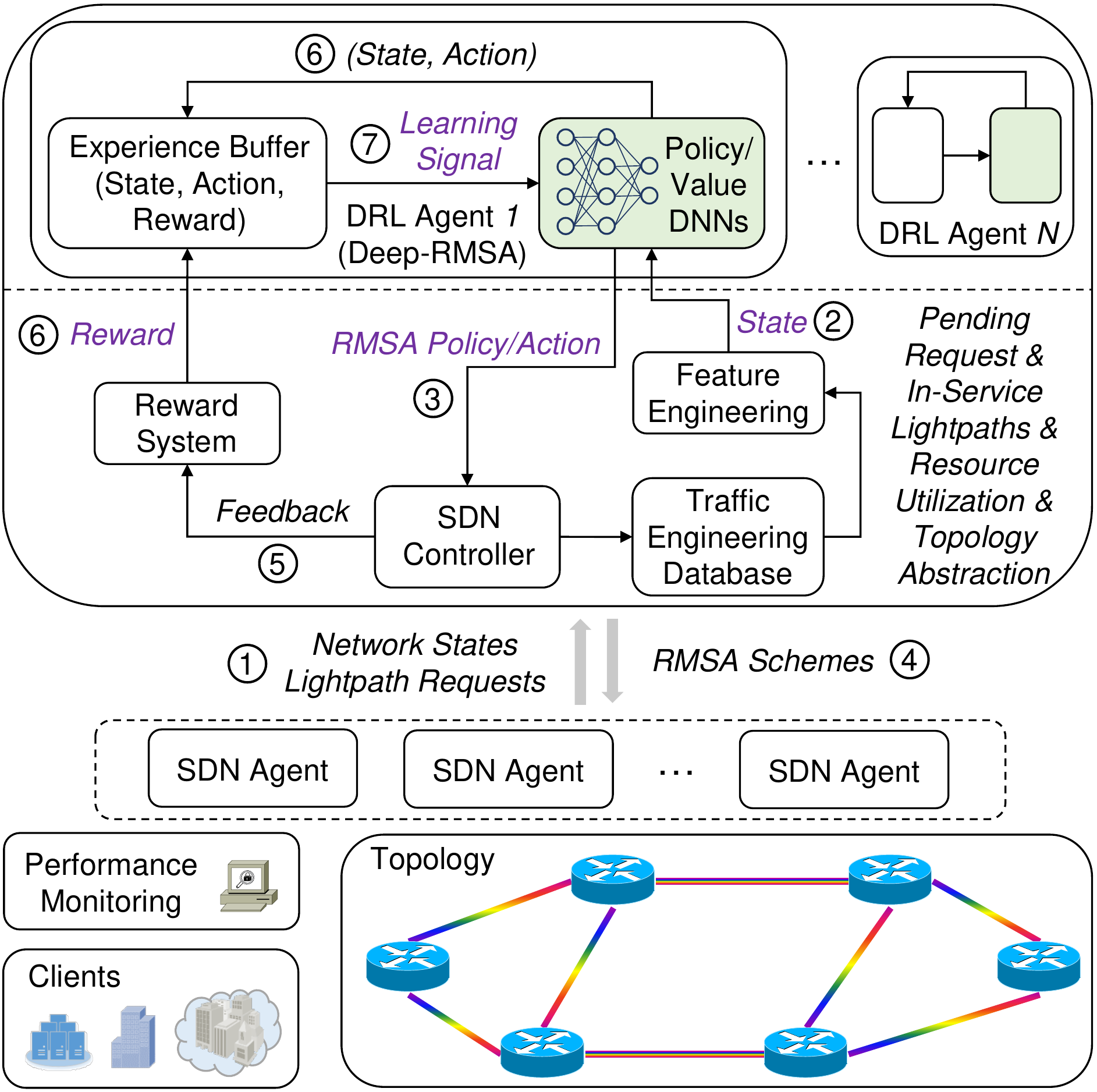} \caption{Schematic of DeepRMSA.}\vspace{-1.2em}
\label{fig:framework}
\end{center}
\end{figure}

Fig.~\ref{fig:framework} shows the schematic of DeepRMSA. DeepRMSA takes advantage of the software-defined networking (SDN) paradigm for centralized and automated control and management of the EON data plane \cite{ChenX2015}. Specifically, a remote SDN controller interacts with the local SDN agents to collect network states and lightpath requests, and distribute RMSA schemes, while the SDN agents drive the actual device configurations according to the received commands. The operation principle of DeepRMSA is designed based on the framework of DRL. Upon receiving a lightpath request $\mathcal{R}_t$ (\emph{step 1}), the SDN controller retrieves from the traffic engineering database key network state representations, including the in-service lightpaths, resource utilization and topology abstraction, and invokes the feature engineering module to generate tailored state data $s_t$ for DeepRMSA (\emph{step 2}). The DNNs of DeepRMSA read the state data and output an RMSA policy $\pi_t$ for the SDN controller (\emph{step 3}). The controller in turn takes an action $a_t$ (i.e., determining an RMSA scheme) based on $\pi_t$ and attempts to set up the corresponding lightpath (\emph{step 4}). The reward system receives the outcome related to the previous RMSA operations as feedback (step 5) and produces an immediate reward $r_t$ for DeepRMSA. $r_t$, together with $s_t$ and $a_t$, are stored in an experience buffer (\emph{step 6}), from which DeepRMSA derives training signals for updating the DNNs afterward (\emph{step 7}). The objective of DeepRMSA upon servicing $\mathcal{R}_t$ is to maximize the long-term cumulative reward defined as,
\begin{equation}\label{eq:2}
\small
\Gamma_t = \sum \limits_{t' \in [t, \infty)} \gamma^{t'-t} \cdot r_{t'},
\end{equation}
where $\gamma \in [0,1]$ is the discount factor that decays future rewards. Eventually, DeepRMSA enables a self-learning capability that can learn and adapt RMSA policies through dynamic lightpath provisioning. Note that, by deploying multiple parallel DRL agents, each for a particular application or functionality (e.g., protection \cite{ChenX2015} and defragmentation \cite{Zhang2016}), we can extend DeepRMSA to build an intact autonomic EON system.

\section{DeepRMSA Design}\label{sec:design}
In this section, we first present the modeling of DeepRMSA, including the definitions of state representation, action space, and reward. Then, we take into account the unique characteristics of dynamic lightpath provisioning and develop two training mechanisms for DeepRMSA.
\subsection{Modeling}\label{sec:model}
\emph{1) State}: The state representation $s_t$ for DeepRMSA is an $1\times(2|V|+1+(2J+3)K)$ array containing the information of $\mathcal{R}_t$ and the spectrum utilization on $K$ candidate paths for $\mathcal{R}_t$. We use $2|V|+1$ elements of $s_t$ to convey $o$, $d$ (in the one-hot format), and $\tau$, where $|V|$ represents the number of nodes in $V$. For each of the $K$ paths, we calculate the sizes and the starting indices of the first $J$ available FS-block, the required number of FS's based on the applicable modulation format, the average size of the available FS-blocks, and the total number of available FS's. Hence, we aim to extract key features on different candidate paths, from which DeepRMSA can sense the global EON state. Note that, a more comprehensive design could include the original two-dimensional spectrum state $F$ in $s_t$ directly to avoid any information loss. However, this would dramatically increase the scale of $s_t$ (i.e., requiring $f_0 \cdot |E|$ elements simply for conveying $F$) and cause scalability issues. Moreover, making DeepRMSA extract useful features from the large-scale binary matrix while incorporating also the topology connectivity and the spectrum continuous and contiguous constraints in EON is not trivial. We will keep this as one of our future research tasks.

\emph{2) Action}: DeepRMSA selects for each $\mathcal{R}_t$ a routing path from the $K$ candidates and one of the $J$ FS-blocks on the selected path. Therefore, the action space (denoted as $A$) includes $K \cdot J$ actions.

\emph{3) Reward}: DeepRMSA receives an immediate reward $r_t$ of $1$ if $\mathcal{R}_t$ is successfully serviced. Otherwise, $r_t = -1$.

\emph{4) DNNs}: DeepRMSA employs a policy DNN $f_{\theta_p}(s_t)$ for generating the RMSA policy (i.e., the probability distribution over the action space) and a value DNN $f_{\theta_v}(s_t)$ for estimating the value of $s_t$ (i.e., the discounted cumulative reward since $s_t$), where $\theta_p$ and $\theta_v$ are the sets of parameters of the DNNs. $f_{\theta_p}(s_t)$ and $f_{\theta_v}(s_t)$ share the same fully-connected DNN architecture \cite{lecun2015} except for the output layers. The output layer of $f_{\theta_p}(s_t)$ consists of $K \cdot J$ neurons, while $f_{\theta_v}(s_t)$ has only one output neuron.

\subsection{Training}
We designed the training of DeepRMSA based on the framework of the A3C algorithm \cite{A3C}. Basically, A3C makes use of multiple parallel actor-learners (child threads of a DRL agent), each interacting with its own copy of the system environment, to achieve learning with more abundant and diversified samples. The actor-learners maintain a set of global DNN parameters $\theta^*_p$ and $\theta^*_v$ asynchronously.

Different from general DRL tasks that can be modeled as Markov decision processes (i.e., the state transition from $s_t$ to $s_{t+1}$ follows a probability distribution given by $P(s_{t+1}|s_t,a_t)$), DeepRMSA involves state transitions which are difficult to be modeled. In particular, due to the fact that $\mathcal{R}_{r+1}$ can be random, there can be infinite possible states for $s_{t+1}$ in DeepRMSA. Thus, we first slightly modified the standard A3C algorithm by defining an episode as the servicing of $N$ lightpath requests, and by making $N$ equal to the training batch size. Here, an episode defines the optimization scope of a DRL task. This way, we eliminate the need for estimating the value of $s_{t+1}$.  We denote DeepRMSA with the episode-based training mechanism as DeepRMSA-EP. \emph{Algorithm} \ref{alg:1} summarizes the procedures of an actor-learner thread in DeepRMSA-EP. In line 1, the actor-learner initiates an empty experience buffer $\Lambda$. Then, for each $\mathcal{R}_t$, the algorithm checks whether $\Lambda$ is empty (i.e., a new episode starts), and if true, synchronizes the local DNNs with the sets of global parameters (lines 3-5). Line 6 updates the EON state by releasing the resources allocated to lightpaths that expire. In line 7, we obtain $s_t$ based on the model discussed in Section \ref{sec:model}. In line 8, we invoke the policy and value DNNs to generate an RMSA policy and a value estimation for $s_t$. Note that, in DeepRMSA-EP, we make $s_t$ include one more element to indicate the position of $\mathcal{R}_t$ regarding the current episode. For instance, if $\mathcal{R}_t$ is the $i$-th request of the episode, we calculate a position indicator as $(N-i+1)/N$. The algorithm decides an RMSA scheme based on the generated policy (lines 9-10, i.e., with the Roulette strategy) and receives a reward accordingly (line 11). The RMSA sample is then stored in the buffer (line 12). With lines 13-21, DeepRMSA-EP performs training every time the buffer contains $N$ samples. Specifically, in the for-loop of lines 14-16, the algorithm first calculates for each sample $\chi_{t'}$ in the buffer the discounted cumulative reward (staring from $\mathcal{R}_{t'}$ till the end of the episode) as,
\begin{equation}\label{eq:3}
\small
\Gamma_{t'} = \sum \limits_{i \in [0,N-1], \chi_{t'+i} \in \Lambda} \gamma^i \cdot r_{t'+i}.
\end{equation}
Then, the advantage of each action being taken can be obtained by,
\begin{equation}\label{eq:4}
\small
\delta_{t'} = \Gamma_{t'} - f_{\theta_v}(s_{t'}),
\end{equation}
which indicates how much an action turns out be better than estimated. Lines 17-18 calculate the policy and values losses $L_{\theta_p}$ and $L_{\theta_v}$, from which policy and value gradients can be derived. In particular, $L_{\theta_p}$ is defined as,
\begin{equation}\label{eq:5}
\small
\begin{aligned}
L_{\theta_p} = &- \frac{1}{N} \sum \limits_{\chi_{t'} \in \Lambda} \delta_{t'} \log f_{\theta_p}(s_{t'},a_{t'})\\
& - \frac{\alpha}{N} \sum \limits_{\chi_{t'} \in \Lambda} \sum \limits_{a \in A} f_{\theta_p}(s_{t'},a) \log f_{\theta_p}(s_{t'},a),
\end{aligned}
\end{equation}
where $\alpha$ ($0<\alpha\ll1$) is a weighting coefficient. The rationale behind Eq~\ref{eq:5} is to reinforce actions (i.e., improving the probabilities) with larger advantages while encouraging exploration (by introducing the total entropy of the policies as a secondary penalty term). The definition of the value loss is straightforward as the mean square error from value estimations, i.e.,
\begin{equation}\label{eq:6}
\small
L_{\theta_v} = \frac{1}{N} \sum \limits_{\chi_{t'} \in \Lambda} \left(f_{\theta_v}(s_{t'}) - \Gamma_{t'}\right)^2.
\end{equation}
Finally, in lines 19-20, the actor-learner applies the gradients to tune the global DNN parameters with training algorithms such as \emph{RMSProp} or \emph{Adam} \cite{Adam}, and empties the buffer to get prepared for the next episode.

\begin{algorithm}
\SetAlgoNoLine
\small
\caption{Procedures of an actor-learner thread in DeepRMSA-EP}\label{alg:1}
initiate $\Lambda=\emptyset$\;
\For{each $\mathcal{R}_t$}
{
    \If{$\Lambda==\emptyset$}
    {
        set $\theta_p = \theta^*_p$, $\theta_v = \theta^*_v$\;
    }
    release the spectra occupied by expired requests\;
    obtain $s_t$ with $\mathcal{R}_t$ and $G(V,E,F)$\;
    calculate $f_{\theta_p}(s_t)$, $f_{\theta_v}(s_t)$\;
    calculate the cumulative sum of $f_{\theta_p}(s_t)$ as $\zeta$\;
    decide an RMSA scheme $a_t = \arg\min \limits_{a} \left\{\zeta(a) \ge rand()\right\}$\;
    attempt to service $\mathcal{R}_t$ with $a_t$ and receive a reward $r_t$\;
    store $(s_t,a_t,f_{\theta_v}(s_t),r_t)$ in $\Lambda$\;
    \If{$|\Lambda|==N$}
    {
        \For{each $\chi_{t'} = (s_{t'},a_{t'},f_{\theta_v}(s_{t'}),r_{t'})$ in $\Lambda$}
        {
            calculate $\Gamma_{t'}$ and $\delta_{t'}$ with Eqs.~\ref{eq:3} and \ref{eq:4}\;
        }
        calculate $L_{\theta_p}$ and $L_{\theta_v}$ with Eqs.~\ref{eq:5} and \ref{eq:6}\;
        obtain the policy and value gradients with $L_{\theta_p}$, $L_{\theta_v}$\;
        apply the gradients to update $\theta^*_p$ and $\theta^*_v$\;
        empty $\Lambda$\;
    }
}
\end{algorithm}%\vspace{-1.5em}

Note that, the uncertainty of dynamic lightpath requests can result in unpredictable trajectories of $s_t$, which in turn can cause oscillations of the cumulative rewards and destabilize the training process. This problem becomes especially severe when the numbers of requests involved are small. Recall the calculation of cumulative rewards in Eq.~\ref{eq:3}, $\Gamma_{t'}$ decreases when $\chi_{t'}$ is getting closer to the end of the buffer and eventually contains the reward from only one request. To cope with this issue, we propose a window-based flexible training mechanism for DeepRMSA, namely DeepRMSA-FLX. Basically, DeepRMSA-FLX invokes the training process each time the buffer contains $2N-1$ samples. DeepRMSA-FLX slides a window of length $N$ through the buffer and calculates the cumulative reward for each of the first $N$ samples, still with Eq.~\ref{eq:3}. Thus, every cumulative reward involves the rewards from servicing $N$ requests. By doing so, we aim to smooth out the oscillations equally for all the samples (if $N$ is sufficiently large\footnote{Note that, we typically set $N$ moderate values, e.g., $50$, to allow training signals being applied to the DNNs quickly.}). Then, the algorithm calculates the policy and value losses with these $N$ samples and updates the global DNN parameters accordingly. The $N$ samples are removed from the buffer afterward. Meanwhile, the condition for synchronizing local DNNs (line 3 of \emph{Algorithm} \ref{alg:1}) becomes $|\Lambda|$ being equal to $N-1$ in DeepRMSA-FLX.

\begin{figure}%[h]
\begin{center}
\includegraphics[width=6cm]{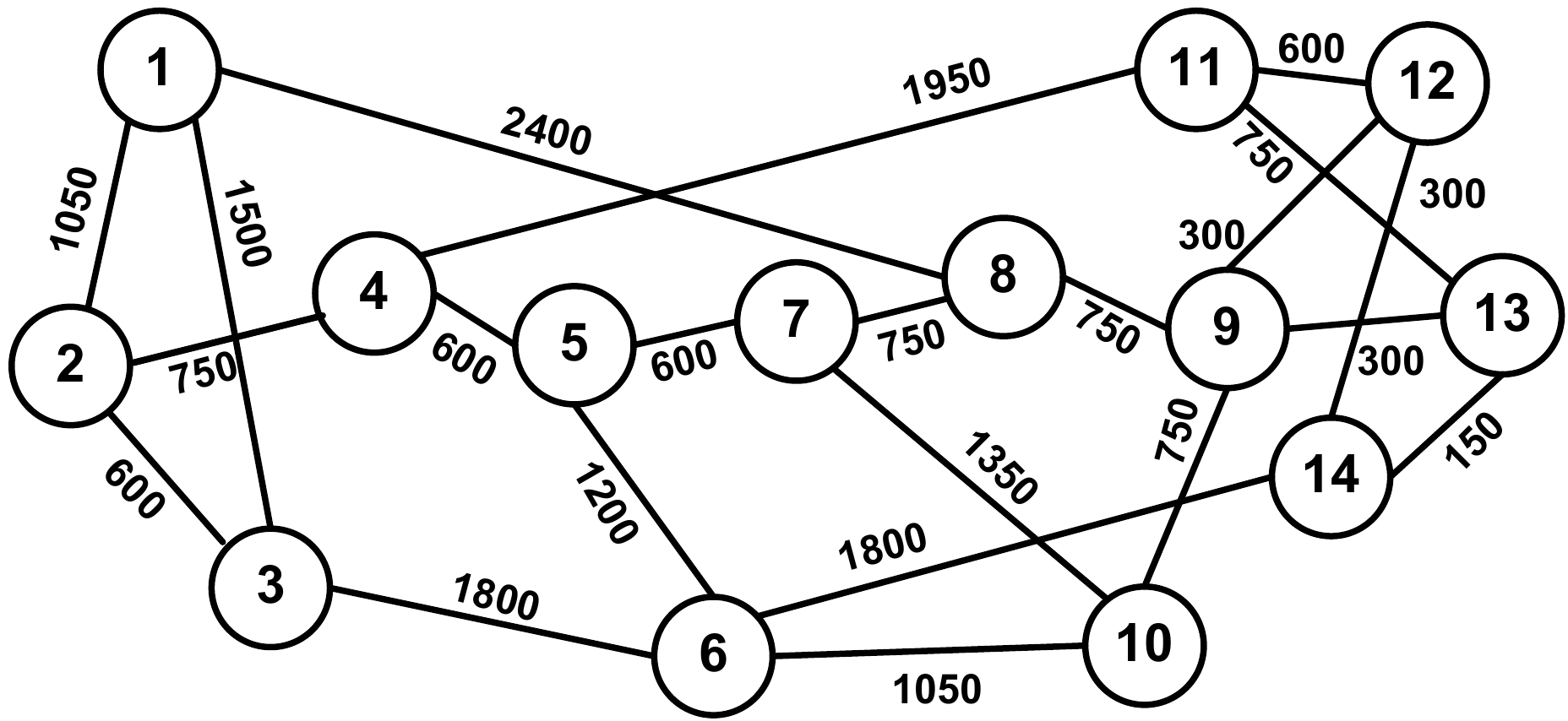} \caption{14-node NSFNET topology (link length in kilometers).}\vspace{-1.2em}
\label{fig:NSF}
\end{center}
\end{figure}

\begin{figure*}%[h!]
\centering
\includegraphics[width=13cm]{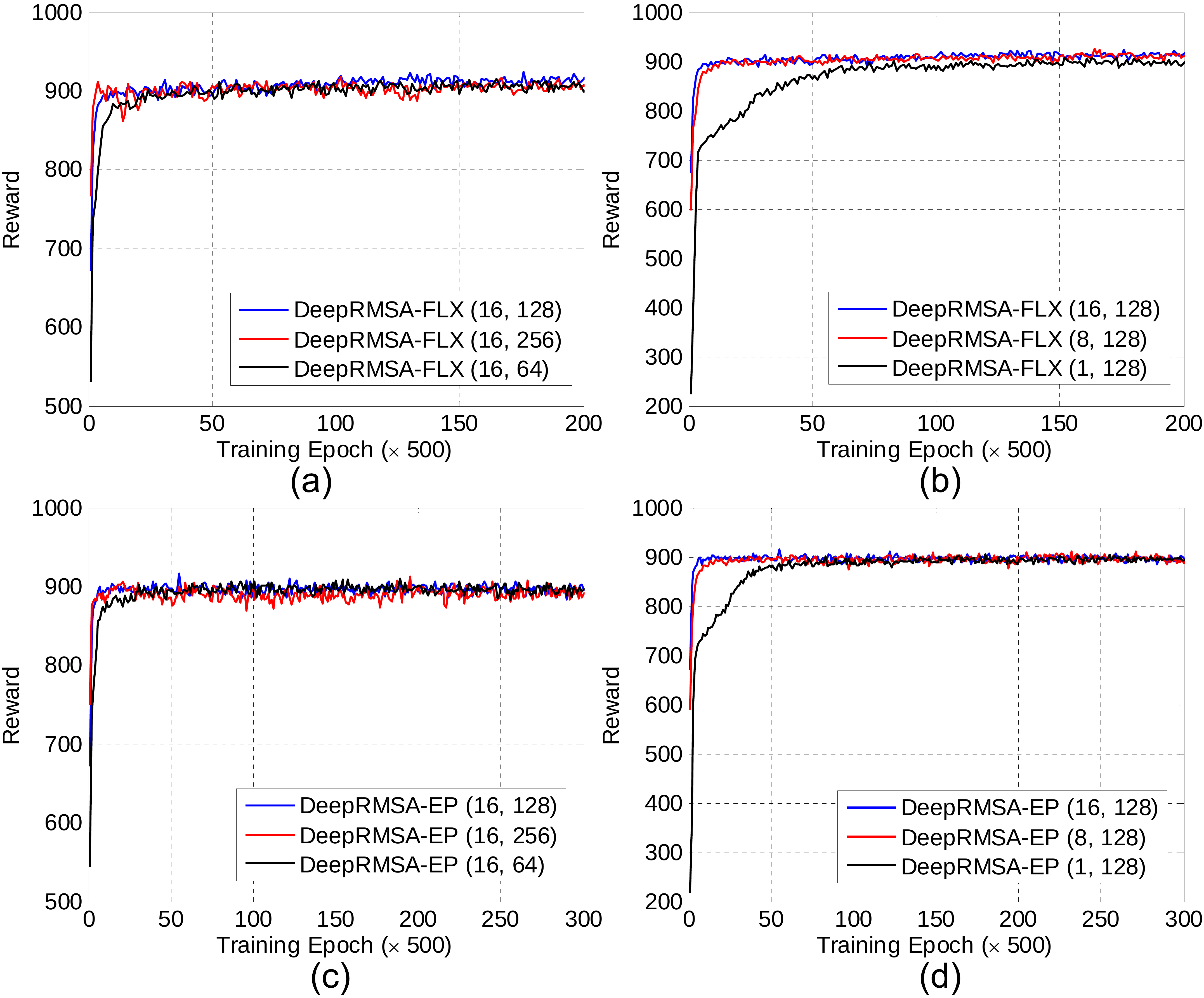}
\caption{Cumulative rewards from DeepRMSA-FLX and DeepRMSA-EP with different (a), (c): DNN sizes, and (b), (d): numbers of actor-learners.}\vspace{-1.2em}
\label{fig:compadnn}
\end{figure*}

\section{Evaluation}\label{sec:evaluation}
\subsection{Simulation Setup}
We evaluated the performance of DeepRMSA with numerical simulations. We first used the 14-node NSFNET topology in Fig.~\ref{fig:NSF} and assumed that each fiber link could accommodate $100$ FS's. The dynamic lightpath requests were generated according to a Poisson process following a uniform traffic distribution, with the average arrival rate and service duration being $10$ and $15$ time units, respectively. The bandwidth requirement of each request is evenly distributed between $25$ and $100$ Gb/s. The DNNs used \emph{ELU} as the activation function for the hidden layers. We set $K=5$ and $J=1$. Hence, DeepRMSA selected only the routing paths and applied the first-fit scheme for spectrum allocation. $\gamma$, $\alpha$, $N$ and the learning rate were set as $0.95$, $0.01$, $50$ and $10^{-5}$, respectively. We used the \emph{Adam} algorithm \cite{Adam} for training. Note that, we normalized every field of $s_t$ before feeding it to the DNNs.

\subsection{Numerical Results}
We first assessed the impact of the scale of the DNNs on the performance of DeepRMSA. We fixed the number of actor-learners as $16$, and implemented DNNs of three setups for both DeepRMSA-EP and DeepRMSA-FLX, i.e., $3$ hidden layers of $64$ neurons ($3 \times 64$), $5$ hidden layers of $128$ neurons ($5 \times 128$), and $8$ hidden layers of $256$ neurons ($8 \times 256$). Figs.~\ref{fig:compadnn}(a) and (c) show the evolutions of cumulative rewards (collected from every $1000$ requests) with different DNN setups during training. We can see that for both of the algorithms, DNNs with larger scales facilitate faster training. In average, it takes DeepRMSA $15,000$ and $5,000$ training epochs to converge with DNNs of $3 \times 64$ and $5 \times 128$ (or $8 \times 256$), respectively. Eventually, the rewards associated with the three setups are very close, with $5 \times 128$ performing slightly better. This is because $5 \times 128$ enables a better ability of data representation when compared with $3 \times 64$, and in the meantime does not suffer from the overfitting issue as encountered by $8 \times 256$. Then, we evaluated the impact of the number of actor-learners by fixing the sizes of the DNNs as $5 \times 128$ and implementing DeepRMSA with $1$, $8$ and $16$ actor-learners. Figs.~\ref{fig:compadnn}(b) and (d) show the corresponding evolutions of cumulative rewards. Again, we can draw the same observations from both of the algorithms, i.e., increasing the number of actor-learners leads to faster convergence and slightly higher rewards. In particular, increasing the number of actor-learners from $1$ to $8$ can accelerate the training speed by a factor of nearly $10$ as multiple parallel actor-learners enable more diversified explorations of the problem. Since the performance gain from further increasing the number of actor-learners is marginal, we expect DeepRMSA with $16$ actor-learners to achieve the best performance. Hence, we fixed the scale of the DNNs and the number of actor-learners as $5 \times 128$ and $16$, respectively, for later evaluations.

\begin{figure}%[h]
\begin{center}
\includegraphics[width=6.5cm]{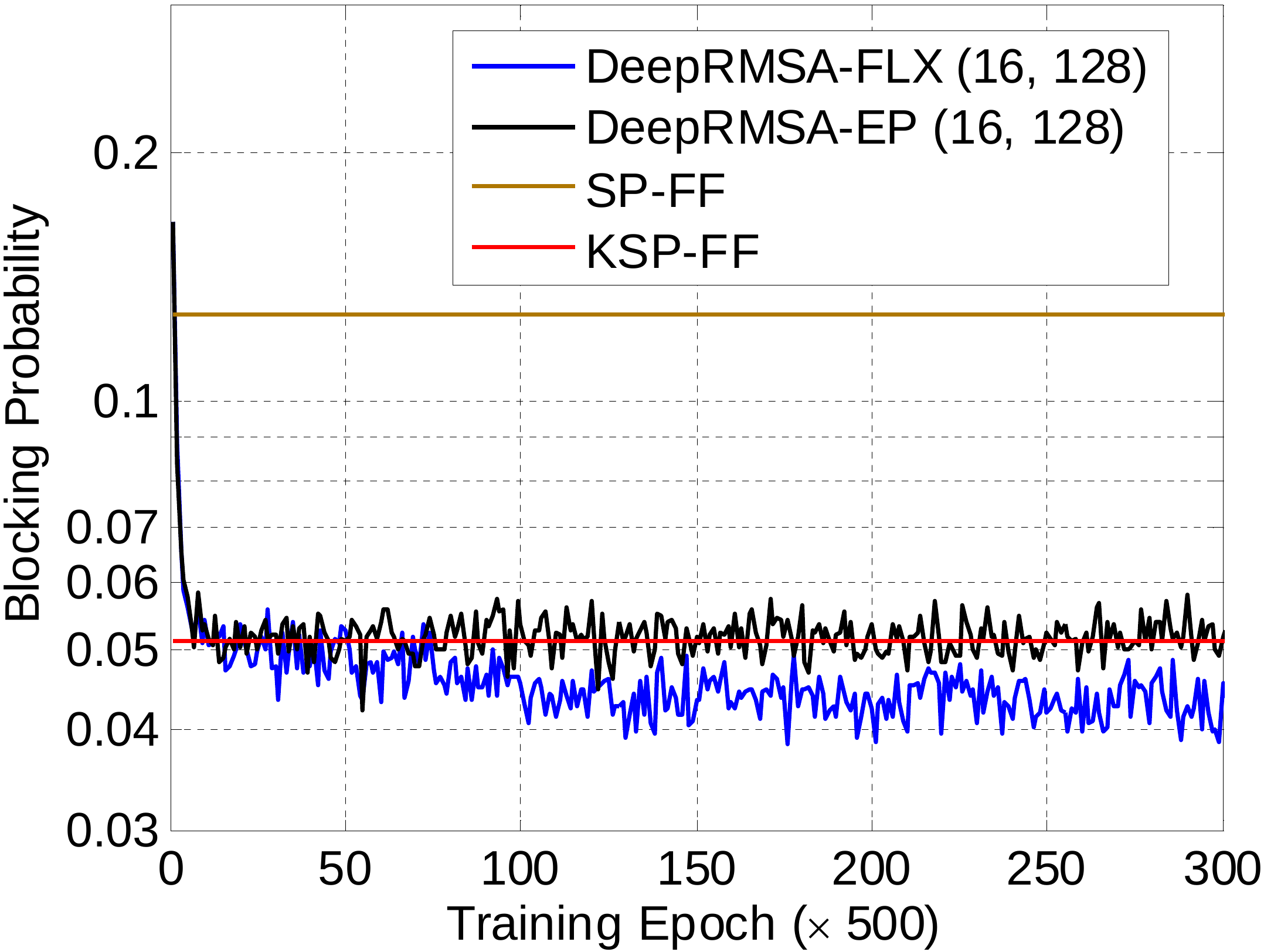} \caption{Request blocking probability.}\vspace{-1.2em}
\label{fig:resultNSF}
\end{center}
\end{figure}

\begin{figure}%[h]
\begin{center}
\includegraphics[width=6.5cm]{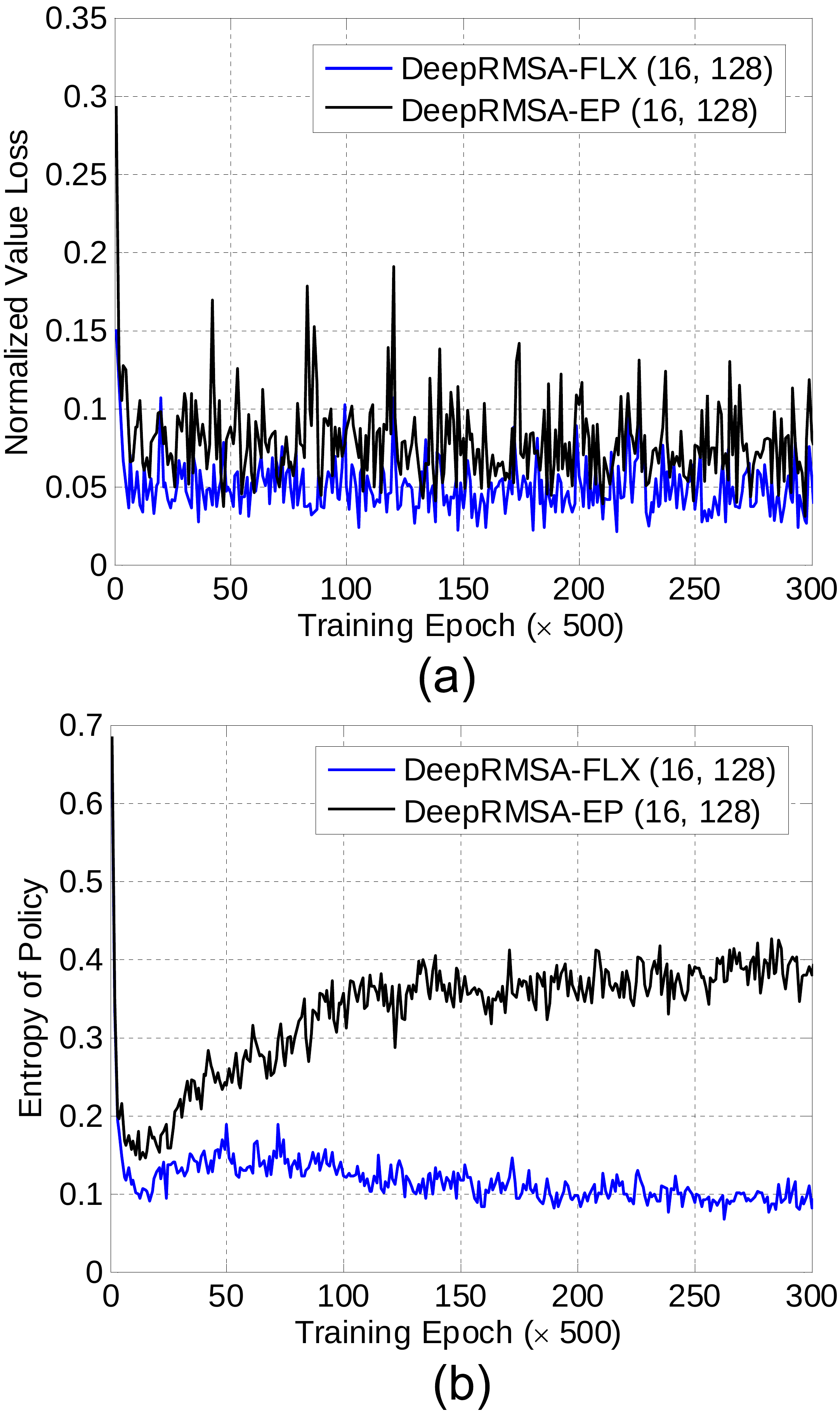} \caption{(a) Normalized value loss, and (b) entropy of policy during training.}\vspace{-1.2em}
\label{fig:valueentropy}
\end{center}
\end{figure}

Next, we compared the performance of DeepRMSA-EP and DeepRMSA-FLX with that of the baseline algorithms, i.e., SP-FF and KSP-FF. KSP-FF has been shown to achieve the state-of-art performance among the existing heuristic designs \cite{Yin2013}. Fig.~\ref{fig:resultNSF} plots the evolution of request blocking probability from the algorithms. We can see that DeepRMSA-EP and DeepRMSA-FLX perform similarly at the beginning and outperform SP-FF after a training period of only $1,000$ epochs. However, DeepRMSA-FLX successfully beats KSP-FF after a training period of $37,500$ epochs, whereas the performance of DeepRMSA-EP eventually merely fluctuates around that of KSP-FF. After training of $150,000$ epochs, DeepRMSA-FLX can achieve a blocking reduction of $20.3\%$ compared with KSP-FF. To reveal the rationale behind the behaviors of DeepRMSA-EP and DeepRMSA-FLX, Figs.~\ref{fig:valueentropy}(a) and (b) present the results of normalized value loss and entropy of policy during training, respectively. It can be seen that the proposed window-based training mechanism facilitates more accurate value estimations (lower value losses) and stabilized training, while the training of DeepRMSA-EP starts to diverge after $10,000$ epochs. Note that, training periods of thousands of epochs are too costly for practical network operations. A more efficient way of training DeepRMSA is expected to be performing offline training with an RMSA simulator first, before enrolling it in online lightpath provisioning for fine tuning \cite{Salman2018}.

\begin{figure}%[h]
\begin{center}
\includegraphics[width=6.5cm]{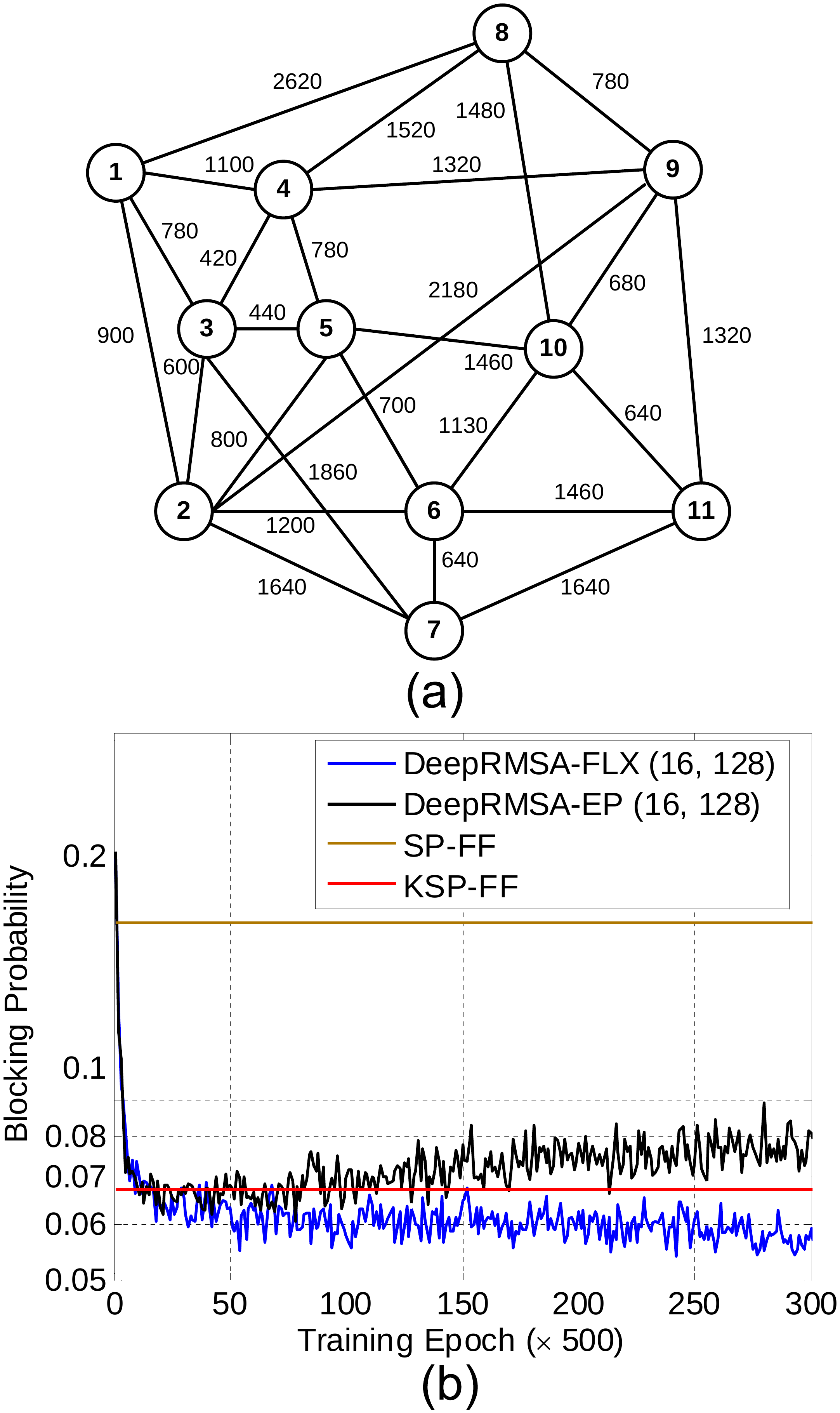} \caption{(a) 11-node COST239 topology (link length in kilometers), and (b) request blocking probability with the COST239 topology.}\vspace{-1.2em}
\label{fig:result239}
\end{center}
\end{figure}

To verify the robustness of DeepRMSA, we also performed simulations with the 11-node COST239 topology in Fig.~\ref{fig:result239}(a). We set the average request arrival rate and service duration as $20$ and $30$ time units, respectively. All the rest of the parameters remained the same as those for the evaluations with the NSFNET topology. Fig.~\ref{fig:result239}(b) shows the results of request blocking probability with the COST239 topology, which demonstrates a clear performance difference between DeepRMSA-EP and DeepRMSA-FLX. Eventually, DeepRMSA-FLX can achieve a blocking probability that is $14.3\%$ and $18.9\%$ lower than those of KSP-FF and DeepRMSA-EP, respectively.

\section{Conclusion}\label{sec:conclusion}
In this paper, we proposed DeepRMSA, a DRL-based RMSA framework for learning the optimal online RMSA policies in EONs. DeepRMSA parameterizes RMSA policies with DNNs and trains the DNNs progressively with experiences from dynamic lightpath provisioning. By taking into account the unique characteristics of the RMSA problem, we developed two training mechanisms for DeepRMSA based on the framework of A3C. Simulation results show that the proposed training mechanisms facilitate successful training of DeepRMSA, which can achieve blocking reductions of more than $20.3\%$ and $14.3\%$ in the NSFNET and COST239 topologies, respectively, when compared with the baselines.

An interesting future research topic would be partitioned DeepRMSA or hierarchical-DeepRMSA where multiple DeepRMSA agents cooperate hierarchically (within the same autonomous system) or interact peer-to-peer through brokers (in a multi-domain EON scenario \cite{multibroker1}) to achieve scalability of DeepRMSA applied to topologies with larger scales. Meanwhile, multi-agent DeepRMSA applied to multiple autonomous system networks will introduce game-theoretic approaches similar to the discussions in \cite{Chen2016broker,ChenJOCN2018}, thus yielding more interesting yet practical multi-agent competitive/cooperative learning problems.

\section*{Acknowledgments}
This work was supported in part by DOE DE-SC0016700, and NSF ICE-T:RC 1836921.

\bibliographystyle{IEEETran}
\bibliography{IEEEabrv,references}

\end{document}